\documentclass[12pt]{iopart}
\usepackage{epsfig}
\usepackage{iopams}  

\newcommand{\beq}{\begin{equation}}
\newcommand{\eeq}{\end{equation}}
\newcommand{\beqa}{\begin{eqnarray}}
\newcommand{\eeqa}{\end{eqnarray}}
\newcommand{\bseq}{\begin{subequations}}
\newcommand{\eseq}{\end{subequations}}

\def\p{{\boldsymbol p}}
\def\0{{\boldsymbol 0}}

\def\simle{\mathrel{\rlap{\raise 0.511ex \hbox{$<$}}{\lower 0.511ex 
\hbox{$\sim$}}}}
\def\simge{\mathrel{ \rlap{\raise 0.511ex 
\hbox{$>$}}{\lower 0.511ex \hbox{$\sim$}}}}

\begin{document}

\title[]{Heavy-flavor dynamics in nucleus--nucleus collisions: from RHIC to LHC}

\author{M. Monteno$^1$, W.M. Alberico$^{2,1}$, A. Beraudo$^{3,4}$, A. De Pace$^1$, \\ 
A. Molinari$^{2,1}$, M. Nardi$^1$ and F. Prino$^1$}

\address{$^1$Istituto Nazionale di Fisica Nucleare, Sezione di Torino, \\
~via P.~Giuria 1, I-10125 Torino, Italy} 
\address{$^2$Dipartimento di Fisica Teorica dell'Universit\`a di Torino, \\ 
~via P.~Giuria 1, I-10125 Torino, Italy} 
\address{$^3$Centro Studi e Ricerche \emph{Enrico Fermi}, Piazza del Viminale 1, Roma, Italy}
\address{$^4$Physics Department, Theory Unit, CERN, CH-1211 Gen\`eve 23, Switzerland}

\ead{monteno@to.infn.it}
\begin{abstract}
The stochastic dynamics of $c$ and $b$ quarks in the fireball created in 
nucleus--nucleus collisions at RHIC and LHC is studied employing 
a relativistic Langevin equation, based on a picture of multiple 
uncorrelated random collisions with the medium. Heavy-quark transport 
coefficients are evaluated within a pQCD approach, with a proper 
HTL resummation of medium effects for soft scatterings. 
The Langevin equation is embedded in a multi-step setup developed to study 
\mbox{heavy-flavor} observables in $pp$ and $AA$ collisions, starting from 
a NLO pQCD calculation of initial heavy-quark yields, complemented in the 
nuclear case by shadowing corrections, \mbox{$k_T$-broadening} and nuclear 
geometry effects. Then, only for $AA$ collisions, the Langevin equation is 
solved numerically in a background medium described by relativistic 
hydrodynamics. Finally, the propagated heavy quarks are made hadronize 
and decay into electrons. Results for the nuclear modification factor 
$R_{AA}$ of heavy-flavor hadrons and electrons from their semi-leptonic decays 
are provided, both for RHIC and LHC beam energies.
\end{abstract}

%\maketitle

Heavy-flavor electron spectra, measured in Au--Au collisions at 
$\sqrt{s}_{NN}=200$~GeV by PHENIX~\cite{PHENIX1,PHENIX2} and 
STAR~\cite{STAR_Erratum} experiments at RHIC, have displayed a large 
suppression with respect to $pp$ collisions, comparable in amount to the 
one observed for charged hadrons. 
Models considering medium-induced gluon radiation~\cite{DjoGyu,ASW}
as the dominant energy loss mechanism for heavy quarks propagating in QGP 
come up against difficulties in reproducing such results. 

These findings gave a great boost to calculations taking in consideration 
the role of collisional energy loss~\cite{vHGR,bamps}. In some models the 
heavy-quark propagation in QGP is described through a Langevin stochastic 
equation~\cite{MooreTeaney,GossiauxAichelin,AkamatsuHirano}, assuming that 
heavy-quark spectrum modifications arise from the cumulated effect of many 
uncorrelated random collisions with the medium.

In our approach we use a relativistic Langevin equation~\cite{Langevin2009}, 
describing the time evolution of the heavy-quark momentum:
\beq\label{eq:lange_r_d}
\frac{\Delta p^i}{\Delta t}=-\eta_D(p)p^i+\xi^i(t)
\eeq
that involves a \emph{deterministic} friction term and a \emph{stochastic} 
noise term $\xi^i(t)$, completely determined by its two-point temporal 
correlator:
\beq\label{eq:noise1}
\fl \langle\xi^i(t)\xi^j(t')\rangle=b^{ij}(\p)\delta(t-t'),\quad{\rm with}\quad
b^{ij}(\p)\equiv \kappa_L(p)\hat{p}^i\hat{p}^j+\kappa_T(p)
(\delta^{ij}-\hat{p}^i\hat{p}^j)
\eeq
The latter involves the transport coefficients 
$\kappa_T(p)\!\equiv\!\frac{1}{2}\frac{\langle \Delta p_T^2\rangle}{\Delta t}$ 
and 
$\kappa_L(p)\!\equiv\!\frac{\langle \Delta p_L^2\rangle}{\Delta t}$, 
which are evaluated according to the procedure presented 
in~\cite{Langevin2011}. We introduce an intermediate 
cutoff $|t|^*\!\sim\!m_D^2$ ($t\!\equiv\!(P'\!-\!P)^2$) to separate hard 
and soft scatterings. 
The contribution of hard collisions ($|t|\!>\!|t|^*$) is evaluated through a 
pQCD calculation of the processes $Q(P)q_{i/\bar i}\!\to\! Q(P')q_{i/\bar i}$ 
and $Q(P)g\!\to\! Q(P')g$. On the other hand for soft collisions 
($|t|\!<\!|t|^*$) a resummation of medium effects is provided by the 
Hard Thermal Loop approximation, with $\alpha_{s}(\mu)$ evaluated at a scale 
$\mu \propto T$. The final result~\cite{Langevin2011} is given by 
$\kappa_{T/L}(p)=\kappa_{T/L}^{\rm hard}(p)+\kappa_{T/L}^{\rm soft}(p)$. 
According to the scale $\mu$ at which $\alpha_{s}(\mu)$ is evaluated to 
calculate $\kappa_{T/L}^{hard}$, we devised two different sets of calculations, 
referred to in the following as HTL1 (for $\mu \propto T$) and HTL2 
(for $\mu=|t|$).
\vspace{0.3cm}

The Langevin simulation tool is embedded in a full setup to calculate 
heavy-flavor observables in $pp$ and $AA$ collisions, divided into the 
following independent steps~\cite{Langevin2011}:
\begin{enumerate}
\item A sample of $c$ and $b$ quarks is generated using 
POWHEG~\cite{POWHEG-hvq}, a code which implements pQCD at NLO accuracy, 
with CTEQ6M PDFs as input. For $AA$ collisions, EPS09 nuclear corrections 
to PDFs are employed~\cite{EPS09}; then, heavy quarks are distributed in 
the transverse plane according to the nuclear overlap function 
$T_{AB}(x,y)\!\equiv\!T_A(x\!+\!b/2,y)T_B(x\!-\!b/2,y)$ 
corresponding to the selected impact parameter $b$; a $p_T$-broadening 
correction to heavy-quark momenta is also included. 
\item At a given proper-time $\tau_{0}$ an iterative 
procedure is started, only for $AA$ collisions, to follow the stochastic 
evolution of the heavy quarks in the plasma until hadronization: 
the Langevin transport coefficients are evaluated at each step  
according to the local 4-velocity and temperature $T(x)$ of the expanding 
background medium, as provided by hydrodynamic codes: both ideal and viscous 
fluid scenarios were tested~\cite{kolb,romatschke1,romatschke2}.
\item Heavy quarks are made fragment into hadron species, sampled 
according to $c$ and $b$ branching fractions taken from 
Refs.~\cite{zeus,PDG2010}, while their momenta are sampled from a Peterson 
fragmentation function~\cite{peterson}, with $\epsilon=0.04$ and $0.005$ for 
$c$ and $b$ respectively. 
Finally, each heavy-quark hadron is forced to decay into electrons with 
PYTHIA~\cite{Pythia}, using updated tables of branching ratios based 
on Ref.~\cite{PDG2010}.
\end{enumerate}

\noindent The effects of the Langevin evolution of $c$ and $b$ quarks in $AA$ 
collisions resulting from our calculations are studied through the 
nuclear modification factor 
$R_{AA}(p_T)\!\equiv\!(dN/dp_T)^{AA}/\langle N\rangle_{\rm coll} (dN/dp_T)^{pp}$ 
and the elliptic flow coefficient 
$v_2(p_T)\!\equiv\!\langle\cos(2\phi)\rangle_{p_T}$ 
of final-state heavy-quark hadrons or decay-electrons. 
Here we will display only results obtained using viscous hydrodynamics for 
some representative values of the input parameters ($\tau_0$ and the QCD scale 
$\mu$), among those fully explored in~\cite{Langevin2011}.

\begin{figure}[t]
\begin{center}
\includegraphics[clip,width=0.40\textwidth]{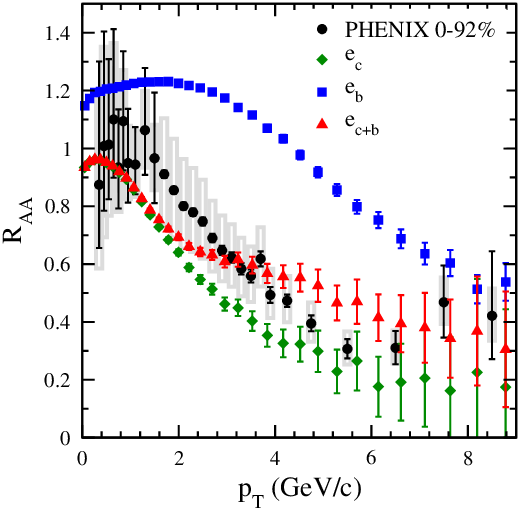}
\includegraphics[clip,width=0.50\textwidth]{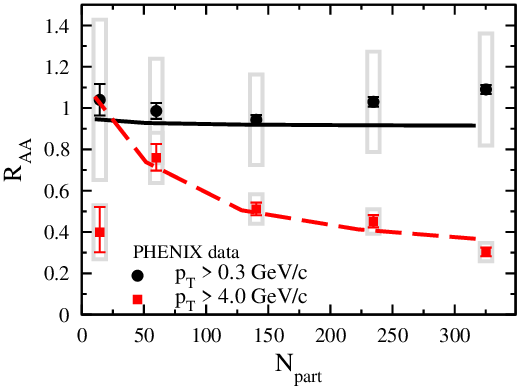}
\caption{(left) $R_{AA}(p_T)$ of heavy-flavor electrons in minimum-bias 
Au--Au collisions at RHIC ($0-92\%$ of the inelastic cross section); 
(right) $R_{AA}$ obtained by integrating the electron yields over the 
indicated momentum ranges, as a function of $N_{part}$. In both panels 
our predictions (with HTL1 calculation) are compared to PHENIX 
data~\cite{PHENIX2}.} 
\label{fig:RAA_PHENIX}
\end{center}
\end{figure}

In Fig.~\ref{fig:RAA_PHENIX} our findings for the $R_{AA}$ of heavy-flavor 
electrons in Au--Au collisions at RHIC ($\sqrt{s}_{NN}=200$~GeV), obtained 
with the calculation HTL1 by assuming viscous hydrodynamics, $\tau_{0}=1$~fm/c 
and $\mu=3 \pi T/2$, are compared to PHENIX~\cite{PHENIX2} data.
In the left panel we observe a general agreement of $R_{AA}(p_{T})$ with the 
PHENIX results on  a minimum-bias data sample ($0-92\%$ of the inelastic cross 
section) for $p_{T} \simge 3$ GeV/c. In the right panel, displaying 
the $R_{AA}$ obtained by integrating the electron yields above a given $p_T$ 
and plotted versus $N_{part}$, the centrality dependence of the 
PHENIX data is shown to be nicely reproduced, except for the most-peripheral 
centrality bin. We do not show here the results obtained for the heavy-flavor 
electron $v_{2}(p_T)$~\cite{Langevin2011}, that appear to underestimate the 
PHENIX minimum-bias data. 
However, a more detailed treatment of hadronization, including also 
the coalescence mechanism, should enhance both $v_2$ and $R_{AA}$ 
at $p_{T} \simle 3$ GeV/c.

\begin{figure}
\begin{center}
\includegraphics[clip,width=0.9\textwidth]{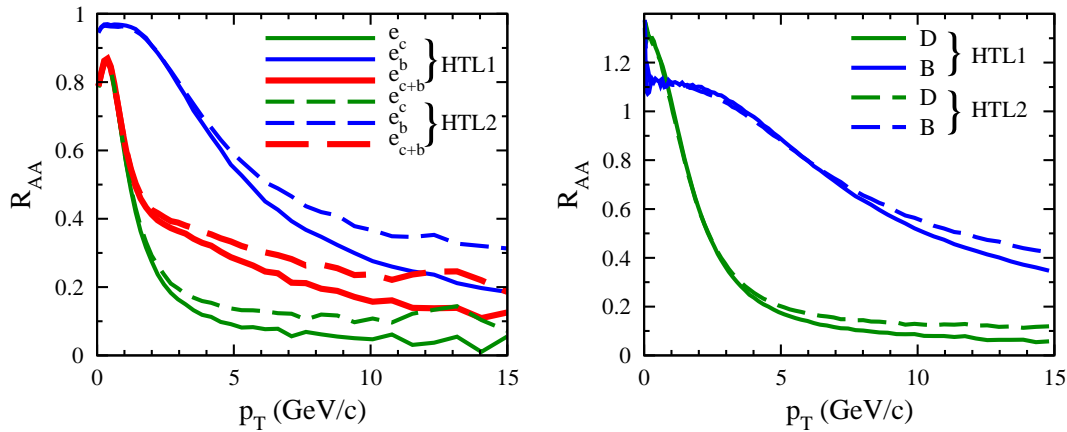}
\vspace{-0.2cm}
\caption{The $R_{AA}$ of heavy-flavor electrons (left) and D, B mesons
(right) in Pb--Pb collisions at LHC ($\sqrt{s}_{NN}=2.76$~TeV) for the 
$0-10\%$ most-central events obtained with the HTL1 (solid line) and 
HTL2 (dashed line) calculations.}
\label{fig:RAA_eA_LHC}
\end{center}
\end{figure}

In Fig.~\ref{fig:RAA_eA_LHC} we show our predictions for the $R_{AA}(p_T)$ of 
heavy-flavor electrons (left panel) and of D, B mesons (right panel) in 
Pb--Pb collisions at LHC ($\sqrt{s}_{NN}=2.76$~TeV), calculated by selecting 
the $0-10\%$ most-central events, under the hypothesis of viscous hydrodynamics
 and for two different choices (HTL1 or HTL2) of the $\mu$ scale in the 
calculation of $\kappa_{T/L}^{hard}$. General features of the $R_{AA}$ 
of heavy-flavor electrons appear similar to those observed at RHIC at the same 
centrality~\cite{Langevin2011}, with a stronger suppression of both charm and 
bottom contributions. However, results obtained with the calculation HTL2 
display a weaker quenching for $p_{T}>3$~GeV/c.
As regards D and B suppression, slighter differences between HTL1 and HTL2 
are observed, and at higher $p_T$.

The main goal of our study was to deliver a weak-coupling calculation to be 
used as a benchmark for advanced studies or less conventional scenarios. 
The capability to accommodate the electron spectra observed at RHIC 
for $p_T\simge 3$~GeV/c strengthens the hypothesis that heavy-quark 
collisional energy loss must be taken into proper account.  
Moreover, we will soon be able to test our predictions against first data with 
Pb--Pb collisions delivered by LHC experiments~\cite{ALICE_e,ALICE_D,CMS_JPSI}.

\vspace{-0.2cm}

\section*{References}
\bibliographystyle{plain}

\end{document}